\documentclass[prb,twocolumn,showpacs,preprintnumbers,amsmath,amssymb]{revtex4}

\usepackage{graphicx}

\begin{document}

\title{All-electron quantum Monte Carlo calculations for the noble gas
atoms He to Xe}

\author{A.~Ma, N.~D.~Drummond, M.~D.~Towler, and R.~J.~Needs}

\affiliation{Theory of Condensed Matter Group, Cavendish Laboratory,
University of Cambridge, Madingley Road, Cambridge, CB3 0HE, United
Kingdom}

\date{\today}

\begin{abstract}
We report all-electron variational and diffusion quantum Monte Carlo
(VMC and DMC) calculations for the noble gas atoms He, Ne, Ar, Kr, and
Xe.  The calculations were performed using Slater-Jastrow wave
functions with Hartree-Fock single-particle orbitals.  The quality of
both the optimized Jastrow factors and the nodal surfaces of the wave
functions declines with increasing atomic number, $Z$, but the DMC
calculations are tractable and well-behaved in all cases.  We discuss
the scaling of the computational cost of DMC calculations with $Z$.
\end{abstract}



\maketitle

\section{Introduction}

The variational quantum Monte Carlo (VMC) method and the more
sophisticated diffusion quantum Monte Carlo (DMC)
method~\cite{foulkes_2001} can yield highly accurate energies for
many-electron systems.  One of the main attractions of these methods
is that the cost of calculating the energy of $N$ quantum particles
scales roughly as $N^{2}$--$N^{3}$, which is better than other
many-body wave function techniques.  However, although the scaling
with particle number is quite advantageous, the cost increases rapidly
with the atomic number $Z$ of the atoms involved.  Theoretical
estimates of this scaling~\cite{hammond_1987,ceperley_1986a} for DMC
calculations have varied from $Z^{5.5}$ to $Z^{6.5}$, while a
practical test~\cite{hammond_1996} indicated a scaling of about
$Z^{5.2}$.

Numerous all-electron DMC studies have been
reported~\cite{reynolds_1982,vrbik_1988,reynolds_1986,lester_1990,
umrigar_1993a,kenny_1995,luechow_1996,huang_1997,
yoshida_1997,shlyakhter_1999,sarsa_2002,casula_2003,casula_2004} for
atoms up to $Z = 10$, but very few have included heavier atoms.  DMC
studies of heavier atoms have normally used pseudopotentials to remove
the chemically inert core electrons from the problem.  However,
pseudopotentials inevitably introduce some errors and it may be useful
to consider how much progress can be made with all-electron DMC
calculations.  Accurate all-electron calculations for atoms may also
be useful in constructing pseudopotentials which incorporate many-body
effects.  In this paper we report VMC and DMC calculations for the
noble gas atoms He, Ne, Ar, Kr, and Xe, which extends the range of
atoms studied within VMC and DMC up to $Z = 54$.  The main aims of
this paper are to investigate how well current all-electron DMC
methods perform for heavy atoms and to study the scaling of the
computational cost with $Z$.

\section{VMC and DMC methods}
\label{sec:qmc}

In VMC the energy is calculated as the expectation value of the
Hamiltonian with an approximate many-body trial wave function
containing a number of variable parameters.  In DMC the estimate of
the ground state energy is improved by performing an evolution of the
wave function in imaginary time.\cite{foulkes_2001} The fermionic
symmetry is maintained by the fixed-node
approximation~\cite{anderson_1976}, in which the nodal surface of the
wave function is constrained to equal that of a trial wave
function. Our DMC algorithm is essentially that of Umrigar \textit{et
al.}~\cite{umrigar_1993a}, and we employ the modifications to the
Green function for all-electron calculations proposed in that paper.
All of our VMC and DMC calculations were performed using the
\textsc{casino} code.\cite{casino}

Our trial wave functions were of the standard Slater-Jastrow form,
\begin{equation}
\label{eq:slater-jastrow}
\Psi = e^{J} D_\uparrow D_\downarrow \ .
\end{equation}
The Jastrow factors, $e^{J}$, were chosen to be functions of the
variables $r_{ij} = |{\bf r}_i - {\bf r}_j|$ and $r_{i} = |{\bf
r}_i|$, where ${\bf r}_i $ is the position of electron $i$ with
respect to the nucleus.  Our Jastrow factors~\cite{drummond_2004} for
He, Ne, Ar, Kr, and Xe contained a total of 26, 75, 79, 80, and 54
adjustable parameters, respectively.\cite{footnote_xe} The optimal
parameter values were obtained by minimizing the variance of the
energy within a VMC procedure.\cite{umrigar_1988,kent_1999} The Slater
determinants, $D_\sigma$, were formed from single-particle orbitals
obtained from Hartree-Fock (HF) calculations using (i) numerical
integration on a radial grid, and (ii) Gaussian basis sets and the
\textsc{crystal98} code.\cite{crystal98} Although the numerical
orbitals are the more accurate they are not available for molecular
systems, in which Gaussian basis sets are very commonly used. 

In both VMC and DMC the energy is calculated as an average over a set
of electron configurations of the local energy, $E_L = \Psi^{-1}
\hat{H} \Psi$, where $\hat{H}$ is the Hamiltonian.  The presence of
core electrons causes two related problems.  The first is that the
shorter length scale variations in the wave function near a nucleus of
large $Z$ require the use of a small timestep.  The second problem is
that the fluctuations in the local energy tend to be large near the
nucleus, because both the kinetic and potential energies are large.
Although these fluctuations can be reduced by optimizing the trial
wave function, in practice they are large for heavier atoms.

At a nucleus the exact wave function has a cusp~\cite{kato_1957} such
that the divergence in the potential energy is canceled by an equal
and opposite divergence in the kinetic energy.  A determinant of exact
HF orbitals obeys the electron-nucleus cusp condition.  However,
Gaussian functions are smooth, and a determinant of such orbitals
cannot have a cusp, so the local energy diverges at the nucleus.  In
practice one finds wild oscillations in the local energy close to the
nucleus, which increase the variance of the energy in VMC calculations
and lead to timestep errors and even numerical instabilities in DMC
calculations.  To solve this problem we make small corrections to the
single-particle orbitals close to the nucleus, which impose the
correct cusp behavior.\cite{cusp_2004}

\section{Results}

\subsection{Quality of the trial wave functions}

The main results of our HF, VMC, and DMC calculations are shown in
Table~\ref{table:total_energies}.  Our HF energy for He with Gaussian
orbitals is very close to the result with the ``exact'' numerical
orbitals, indicating the high quality of the Gaussian basis set used.
For Ne the HF energy with Gaussian orbitals is a little higher than
the value with the numerical orbitals, although this difference is not
large enough to affect the DMC results.  We experimented with various
Gaussian basis sets for the heavier noble gas atoms and found that the
basis set errors at the HF level tend to increase significantly with
atomic number. In the case of Xe our best Gaussian basis set gave an
error of 0.11~a.u.  For Ar, Kr, and Xe we therefore used only the
numerical orbitals.

The exact ground-state wave function of a two-electron atom is a
nodeless function of $r_1$, $r_2$, and $r_{12}$, which is the same
form as our trial wave function for He.  We therefore expect to obtain
a highly accurate trial wave function for He.  We refer to the
difference in the HF and DMC energies as the ``DMC correlation
energy''.  If one keeps the orbitals fixed and varies the Jastrow
factor, then the lowest energy one could obtain is the DMC energy.
The percentage of the DMC correlation energy retrieved at the VMC
level is therefore a measure of the quality of the Jastrow factor.
From the data in Table~\ref{table:total_energies} we find that our VMC
calculations retrieve 99.5\%, 91\%, 85\%, 70\%, and 59\% of the DMC
correlation energy for He, Ne, Ar, Kr, and Xe, respectively.  We
believe that the decrease in the quality of the Jastrow factor with
increasing $Z$ arises from the increasing inhomogeneity of the atoms.
Creating accurate Jastrow factors for all-electron studies of heavy
atoms is a challenging problem.

\begin{table}[t]
\begin{tabular}{c c c l c }

\hline
\hline
{\bf Atom}& {\bf Method} & {\bf Orb.} & {\bf Total energy} & $E_{\rm c}$ \\
        &        & {\bf type} & {\bf (a.u.)}  &    \\
\hline
   & HF  & G & $-2.86165214$  &  0 \% \\
   & HF  & N & $-2.86168000$  &  0 \% \\ 
   & VMC & G & $-2.903499(8)$ & 99.5 \% \\ 
He & VMC & N & $-2.903527(9)$ & 99.5 \% \\ 
   & DMC & G & $-2.903732(5)$ & 100 \% \\ 
   & DMC & N & $-2.903719(2)$ & 100 \% \\
   & ``Exact''~\cite{pekeris_58} & - & $-2.903724$ & 100 \% \\ 
\hline
   & HF  & G & $-128.53832860$  &  0 \%  \\ 
   & HF  & N & $-128.54709811$  &  0 \%  \\  
   & VMC & G & $-128.8794(4)$   & 85 \%  \\ 
Ne & VMC & N & $-128.891(5)$    & 88 \%  \\ 
   & DMC & G & $-128.9232(5)$   & 96 \%  \\ 
   & DMC & N & $-128.9231(1)$   & 96 \%  \\
   & ``Exact''~\cite{davidson_91} & - & $-128.939$ & 100 \% \\ 
\hline
   & HF  & N & $-526.81751277$ &   0 \% \\  
Ar & VMC & N & $-527.3817(2)$  &  77 \% \\ 
   & DMC & N & $-527.4840(2)$  &  91 \% \\
   & ``Exact''~\cite{davidson_1996} & - & $-527.55$ & 100 \% \\  
\hline 
   & HF  & N & $-2752.05497715$ &  0 \% \\  
Kr & VMC & N & $-2753.2436(6)$  & 57 \% \\ 
   & DMC & N & $-2753.7427(6)$  & 82 \% \\
   & ``Exact''~\cite{clementi_1995} & - & $-2754.13$ & 100 \% \\ 
\hline 
   & HF  & N & $-7232.13836331$ &  0 \% \\  
Xe & VMC & N & $-7233.700(2)$   & 46 \% \\  
   & DMC & N & $-7234.785(1)$   & 77 \% \\
   & ``Exact''~\cite{clementi_1995} & - & $-7235.57$ & 100 \% \\ 
\hline
\hline
\end{tabular}
\caption{Total energies of the noble gas atoms, and the percentages of
the correlation energies, $E_{\rm c}$, retrieved.  (G) denotes a
calculation with a Gaussian basis set and (N) denotes numerical
orbitals.  The ``exact'' energies were obtained from data in the
indicated references.}
\label{table:total_energies}
\end{table}


Our VMC and DMC energies for Ne obtained with the numerical orbitals
are very close to those obtained in our earlier
work.\cite{drummond_2004}  Huang \textit{et al.}~\cite{huang_1997}
obtained a VMC energy of $-128.9008(1)$~a.u., which is only slightly
lower than our value, although they also optimized the orbitals. Our
DMC energies for Ne are within error bars of those reported by Umrigar
\textit{et al.}~\cite{umrigar_1993a}, but the remaining fixed-node
error of 0.016~a.u. is substantial.

From the data in Table~\ref{table:total_energies} we observe that the
percentages of the correlation energy missing at the DMC level are
0\%, 4\%, 9\%, 18\% and 23\% for He, Ne, Ar, Kr and Xe, respectively.
This indicates that the size of the fixed-node error increases rapidly
with $Z$.

\subsection{Theoretical scaling with atomic number}
\label{subsec:theory}

It is of interest to study the CPU time required to obtain a fixed
standard error in the mean energy, $\Delta$, as a function of the
atomic number $Z$. The required CPU time, $T$, can be written as
\begin{equation}
\label{eq:T}
T \propto M \, T_C \;,
\end{equation}
where $M$ is the total number of generations of electron
configurations and $T_C$ is the CPU time for one move of $C$
configurations, where $C$ is the average number of configurations in a
generation.

Ceperley~\cite{ceperley_1986a} showed that $\Delta^2$ can be written
as the sum of two terms; the first corresponds to the square of the
standard error evaluated as if the DMC energies were uncorrelated, and
the second accounts for the effects of correlations.  In DMC
calculations the timestep $\tau$ is normally chosen to be small, and
the correlations between configurations at successive generations are
large, so that the second of these terms dominates.  Ceperley showed
that this term is given approximately by~\cite{ceperley_1986a}
\begin{equation}
\label{eq:ceperley_sigma^2}
\Delta^2 = \frac{2|E_{\rm VMC} - E_{\rm DMC}|}{\tau M C} \;.
\end{equation}
Since $\Delta^2$ is inversely proportional to the total number of
configurations, we obtain
\begin{equation}
\label{eq:T_final}
T \propto \tau^{-1} |E_{\rm VMC} - E_{\rm DMC}| \, T_C \;.
\end{equation}

Ceperley used the simple approximation $|E_{\rm VMC} - E_{\rm DMC}|
\propto E_{\rm c}$, and the approximate scaling $E_{\rm c} \propto
Z^{1.5}$.  He also argued that avoiding large timestep errors requires
$\tau \propto Z^{-2}$, as the average distance diffused should be
smaller than the size of the $1s$ orbital, which is proportional to
$Z^{-1}$.  Finally, he used $T_C \propto Z^2$ to obtain an overall
scaling of
\begin{equation}
T \propto Z^{5.5} \;.
\end{equation}
Hammond \textit{et al.}~\cite{hammond_1987} argued along similar
lines, although they chose $T_C \propto Z^3$, leading to an overall
scaling of $T \propto Z^{6.5}$.  In what follows we examine some
aspects of these arguments.

\subsection{Numerical tests of scaling with atomic number}

Discussions of the actual scaling of the computational cost of
calculations with system size or atomic number are fraught with
difficulties.  The results depend on the computers on which the
calculations are run, the algorithms used, and on the details of the
software used.  Our calculations are run on parallel computers in
which each processor deals with a small number of electronic
configurations (one in VMC and roughly ten in DMC).  The
interprocessor communications are negligible in VMC and small in DMC,
and the computational cost is inversely proportional to the number of
processors used.  All of the DMC results used for determining the
scaling of the computational cost with atomic number were performed on
96 processors of a Sunfire Galaxy machine, although most of the
variance minimizations were performed on a cluster of 16 Xeon
dual-processors.

To ensure that timestep errors are small the DMC timestep should be
chosen so that the probability of a move being accepted is high.  For
the DMC results reported in Table~\ref{table:total_energies} we used
timesteps of 0.02, 0.0025, 0.0009, 0.00035, and 0.0002~a.u., for He,
Ne, Ar, Kr, and Xe, respectively, which were chosen so that in each
case slightly more than 99 \% of the proposed moves were accepted.
These timesteps scale as $Z^{-1.41}$, which is significantly weaker
than the $Z^{-2}$ scaling used in the earlier theoretical
estimates.\cite{hammond_1987,ceperley_1986a} We therefore expect the
timestep bias in our DMC results to increase with $Z$.

We thoroughly investigated the timestep dependence of the energies for
He and Ne, concluding that they are negligible compared with the
statistical error bars given in Table~\ref{table:total_energies}.  For
each of Ar, Kr and Xe we performed calculations at four different
timesteps, and we estimate that the timestep errors in the
corresponding DMC energies are less than 0.002~a.u.\ (Ar), 0.01~a.u.\
(Kr) and 0.015~a.u.\ (Xe).  It is likely that the larger timestep
errors in our DMC results for the heavier atoms arise both from the
reduction in the quality of the trial wave functions and the poorer
sampling of the core electrons.

The correlation energy is normally defined as the difference between
the exact non-relativistic ground state energy and the HF energy,
assuming static point nuclei.  Accurate estimates of the correlation
energies of neutral atoms for $Z=2$ to $18$ are given by Davidson and
Chakravorty~\cite{davidson_1996}, while Clementi and
Hofmann~\cite{clementi_1995} give values for Kr and Xe which, while
probably not as accurate as those for the lighter atoms, are expected
to be quite reliable.  We will take these as our reference data and
refer to them as the ``exact'' correlation energies, and when added to
the Hartree-Fock energies, the ``exact'' energies.

\begin{figure}[t]
\begin{center}
\includegraphics[width= \columnwidth]{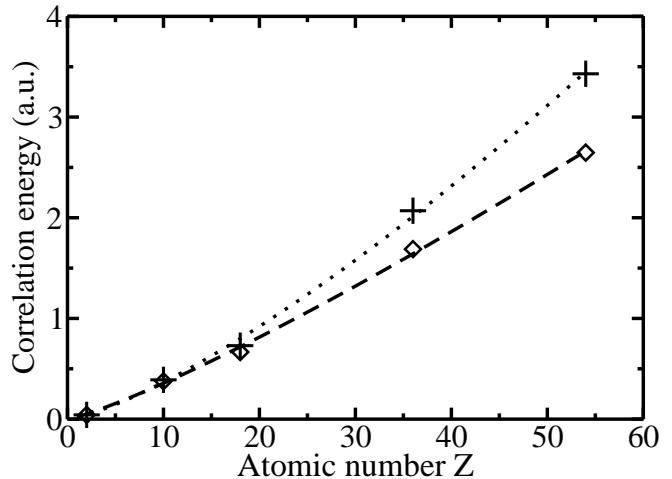}
\end{center}
\caption{The correlation energy $E_{\rm c}$ as a function of atomic
number $Z$. Crosses: ``exact'' values, diamonds: DMC values.  The
dotted line is a fit to the ``exact'' values giving $E_{\rm c} \propto
Z^{1.33}$, while the dashed line is a fit to the DMC data giving
$E_{\rm c} \propto Z^{1.26}$.}
\label{fig:corr_e}
\end{figure}

Fig.~\ref{fig:corr_e} shows the correlation energy as a function of
$Z$ from our DMC data and the estimates of Davidson and
Chakravorty~\cite{davidson_1996} ($Z$=2--18), and Clementi and
Hofmann~\cite{clementi_1995} ($Z=36,54$).  It is clear that (apart
from He) DMC underestimates the correlation energy, and that the
underestimation becomes more severe at larger $Z$.  The best power-law
fit to the ``exact'' data for the noble gas atoms gives $E_{\rm c}
\propto Z^{1.33}$, while for our DMC data we obtain $Z^{1.26}$.  The
scaling of $Z^{1.5}$ assumed in the earlier theoretical
estimates~\cite{hammond_1987,ceperley_1986a} is somewhat of an
overestimate.

As mentioned in Sec.~\ref{subsec:theory}, the quantity which actually
enters Ceperley's approximation of Eq.~(\ref{eq:ceperley_sigma^2}) for
the variance of the DMC energy is the difference between the
variational and DMC energies, $|E_{\rm VMC} - E_{\rm DMC}|$. Using the
VMC and DMC data given in Table~\ref{table:total_energies} we find
$|E_{\rm VMC} - E_{\rm DMC}| \propto Z^{2.62}$.  The reason that this
quantity increases more rapidly with $Z$ than $E_{\rm c}$ is that the
percentage of the correlation energy retrieved in our VMC calculations
decreases with $Z$ more rapidly than in our DMC calculations.

We can also test Eq.~(\ref{eq:ceperley_sigma^2}) directly by comparing
the difference between the variational and DMC energies $|E_{\rm VMC}
- E_{\rm DMC}|$ with the quantity $ \frac{1}{2}\tau MC
\Delta^2$. Serial correlation of the data has been taken into account
when computing the variance over the run.  The results shown in
Table~\ref{fig:variances} indicate that the two quantities are in good
agreement, which is rather satisfactory considering the large range of
$Z$ and the very different qualities of trial wave functions used.
The quantity $\frac{1}{2}\tau M C \Delta^2$ is fitted by a scaling of
$Z^{2.71}$.

\begin{table}[t]
\begin{tabular}{ccc}
\hline
\hline
Atom & $ \frac{1}{2} \tau M C \Delta^2$ (a.u.) & $|E_{\rm VMC}-E_{\rm DMC}|$ (a.u.)\\
\hline
He & 0.00019 & 0.00019 \\
Ne & 0.024   & 0.032 \\
Ar & 0.11    & 0.10 \\
Kr & 0.66    & 0.50 \\
Xe & 1.3     & 1.0 \\
\hline
\hline
\end{tabular}
\caption{The quantity $\frac{1}{2} \tau M C \Delta^2$ and the
difference between the VMC and DMC energies.}
\label{fig:variances}
\end{table}


\begin{figure}[t]
\begin{center}
\includegraphics[width= \columnwidth]{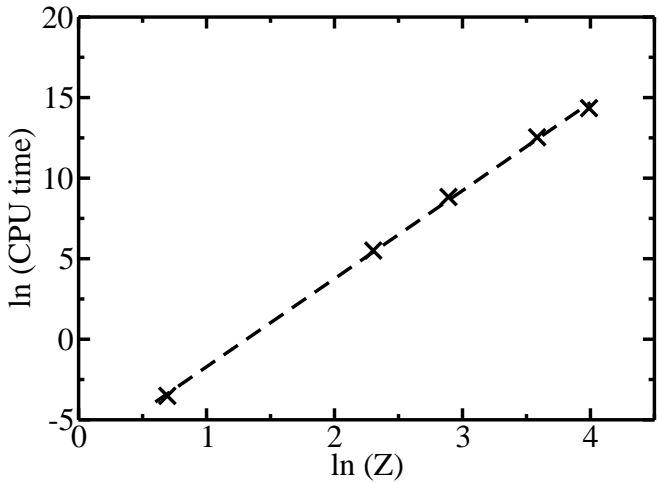}
\end{center}
\caption{The logarithm of the CPU time required to obtain a fixed
error bar in the energy versus $\ln(Z)$ for our DMC calculations.  The
dashed line shows the fitted scaling of $Z^{5.47}$. The CPU times
are measured in seconds.}
\label{fig:scaling}
\end{figure}

We found that the computational cost of moving all the electrons in a
configuration scaled as $Z^{1.35}$ in our DMC calculations.  This is
rather better than the scalings assumed by
Ceperley~\cite{ceperley_1986a} ($Z^2$) and by Hammond \textit{et
al.}~\cite{hammond_1987} ($Z^3$).  If we studied a system containing
many atoms, the scaling of the computational cost for moving all the
electrons in a configuration would be expected to increase roughly as
$N^{2}$, although the use of localized Wannier functions could reduce
this to $N$.\cite{williamson_2001}

Putting together our scalings for the factors in
Eq.~(\ref{eq:T_final}) we find
\begin{equation}
\label{eq:T(Z)}
T \propto Z^{1.41} \times Z^{2.62} \times Z^{1.35} = Z^{5.38}\;.
\end{equation}
We can now compare this with the actual DMC computations reported in
Table~\ref{table:total_energies}.  In Fig.~\ref{fig:scaling} we show
the logarithm of the CPU time as a function of $Z$ for 
a given standard error of the mean.
The best fit gives a scaling of $Z^{5.47}$, in good agreement
with the prediction of $Z^{5.38}$ from Eq.~(\ref{eq:T_final}).


As mentioned before, our DMC results for the heavier atoms suffer from
significant timestep errors.  If we adopt the $Z^{-2}$ scaling for the
timestep instead of the $Z^{-1.41}$ used above, we obtain an overall
scaling of $T \propto Z^{5.97}$, which is higher than the value of $T
\propto Z^{5.2}$ obtained in the practical tests of Hammond \textit{et
al.}\cite{hammond_1996} Moreover, it seems likely that an even more
rapid scaling would be required to achieve a timestep error
independent of $Z$.

\section{Conclusions}

We have applied the VMC and DMC methods to noble gas atoms up to Xe
($Z = 54$), using Slater-Jastrow wave functions with Hartree-Fock
single-particle orbitals.  The percentage of the DMC correlation
energy obtained at the VMC level decreases with $Z$, indicating that
the quality of our Jastrow factors decreases with $Z$.  The percentage
of the exact correlation energy retrieved at the DMC level also
decreases with $Z$, indicating that the quality of the HF nodal
surface deteriorates with increasing $Z$.

Our study shows that Ceperley's expression~\cite{ceperley_1986a} for
the variance of the DMC energy (Eq.~(\ref{eq:ceperley_sigma^2})) is
accurate to better than a factor of two for the systems studied here.
The computational cost required to obtain a fixed statistical error
bar in the energy scaled as $Z^{5.47}$, but in these calculations the
timestep error increased significantly with $Z$.  The scaling required
to achieve a timestep error independent of $Z$ is difficult to
estimate, but it would certainly be higher than $Z^{5.47}$.  However,
it may well be reasonable to incur substantial timestep errors deep in
the core of the atom when we calculate chemical properties which are
related to the valence electrons.

\section{Acknowledgments}
 
We thank John Trail for providing the numerical Hartree-Fock atomic
orbitals and energies.  We acknowledge financial support from the
Engineering and Physical Sciences Research Council of the United
Kingdom. Access to the Sunfire Galaxy computer was provided by the
Cambridge-Cranfield High Performance Computing Facility.


\begin{thebibliography}{}

\bibitem{foulkes_2001} W.~M.~C.~Foulkes, L.~Mitas, R.~J.~Needs, and
G.~Rajagopal, Rev.~Mod.~Phys. \textbf{73}, 33 (2001).

\bibitem{hammond_1987} B.~L. Hammond, P.~J. Reynolds, and
W.~A. Lester, Jr., J. Chem. Phys. \textbf{87}, 1130 (1987).

\bibitem{ceperley_1986a} D.~M. Ceperley, J. Stat. Phys. \textbf{43},
  815 (1986).

\bibitem{hammond_1996} Data from B.~L. Hammond, as reported in
D.~M. Ceperley and L.~Mitas, Adv. Chem. Phys. \textbf{93}, 1 (1996).

\bibitem{reynolds_1982} P.~J.~Reynolds, D.~M.~Ceperley, B.~J.~Alder,
and W.~A.~Lester, Jr., J. Chem. Phys. \textbf{77}, 5593 (1982).

\bibitem{reynolds_1986} P.~J.~Reynolds, R.~N.~Barnett, B.~L.~Hammond,
and W.~A.~Lester, J. Stat. Phys. \textbf{43}, 1017 (1986).

\bibitem{vrbik_1988}J.~Vrbik, M.~F.~DePasquale, and S.~M.~Rothstein,
J. Chem. Phys. \textbf{88}, 3784 (1988).

\bibitem{lester_1990} W.~A.~Lester, Jr., and B.~L.~Hammond,
Ann. Rev. Phys. Chem. \textbf{41}, 283 (1990).

\bibitem{umrigar_1993a} C.~J.~Umrigar, M.~P.~Nightingale, and
K.~J.~Runge, J.~Chem.~Phys. \textbf{99}, 2865 (1993).

\bibitem{kenny_1995} S.~D.~Kenny, G.~Rajagopal, and R.~J.~Needs,
Phys. Rev. A \textbf{51}, 1898 (1995).

\bibitem{luechow_1996} A.~L{\"u}chow and J.~B.~Anderson,
J. Chem. Phys. \textbf{105}, 7573 (1996).

\bibitem{yoshida_1997} T.~Yoshida and G.~Miyako,
J. Chem. Phys. \textbf{107}, 3864 (1997).

\bibitem{huang_1997} C.~J.~Huang, C.~J.~Umrigar, and M.~P.~Nightingale,
J. Chem. Phys. \textbf{107}, 3007 (1997).

\bibitem{shlyakhter_1999} Y.~Shlyakhter, S.~Sokolova, A.~L{\"u}chow, 
and J.~B.~Anderson, J. Chem. Phys. \textbf{110}, 10725 (1999).

\bibitem{sarsa_2002} A.~Sarsa, J.~Boronat, and J.~Casulleras,
J. Chem. Phys. \textbf{116}, 5956 (2002).

\bibitem{casula_2003} M.~Casula and S.~Sorella,
J. Chem. Phys. \textbf{119}, 6500 (2003).

\bibitem{casula_2004} M. Casula, C. Attaccalite, and S. Sorella,
J. Chem. Phys. \textbf{121}, 7110 (2004).

\bibitem{anderson_1976} J.~B.~Anderson, J. Chem. Phys. \textbf{65},
  4121 (1976).

\bibitem{casino} R.~J.~Needs, M.~D.~Towler, N.~D.~Drummond, and
P.~R.~C.~Kent, \textsc{casino} version 1.7 User Manual, University of
Cambridge, Cambridge (2003).

\bibitem{drummond_2004} N.~D.~Drummond, M.~D.~Towler, and R.~J.~Needs,
Phys. Rev. B \textbf{70}, 235119 (2004).

\bibitem{umrigar_1988} C.~J.~Umrigar, K.~G.~Wilson, and J.~W.~Wilkins,
Phys. Rev. Lett. \textbf{60}, 1719 (1988).

\bibitem{kent_1999} P.~R.~C.~Kent, R.~J.~Needs and G.~Rajagopal,
Phys.~Rev.~B \textbf{59}, 12344 (1999).

\bibitem{footnote_xe} For Xe, VMC energies of $-7233.700(2)$~a.u. and
$-7233.713(4)$~a.u. were obtained using Jastrow factors with 54 and 80
parameters, respectively.  Since the lowering of the energy due to the
use of a larger number of parameters was small, the 54-parameter
Jastrow factor was used in the DMC calculations.

\bibitem{crystal98} V.~R.~Saunders, R.~Dovesi, C.~Roetti, M.~Caus\`a,
N.~M.~Harrison, R.~Orlando, and C.~M.~Zicovich-Wilson,
\textsc{crystal98} \textit{User's Manual}, University of Torino,
Torino (1998).

\bibitem{kato_1957} T. Kato, Commun.~Pure~Appl.~Math. \textbf{10}, 151
  (1957).

\bibitem{cusp_2004} A.~Ma, M.~D.~Towler, N.~D.~Drummond, and
R.~J.~Needs, unpublished.

\bibitem{pekeris_58} C.~L.~Pekeris, Phys. Rev. \textbf{112}, 1649
  (1958).

\bibitem{davidson_91} E.~R.~Davidson, S.~A.~Hagstrom,
S.~J.~Chakravorty, V.~M.~Umar, and C.~F.~Fischer, Phys. Rev. A
\textbf{44}, 7071 (1991); S.~J.~Chakravorty, S.~R.~Gwaltney,
E.~R.~Davidson, F.~A.~Parpia, and C.~F.~Fischer, Phys. Rev. A
\textbf{47}, 3649 (1993).

\bibitem{davidson_1996} S.~J.~Chakravorty and E.~R.~Davidson,
J. Phys. Chem. \textbf{100}, 6167 (1996).

\bibitem{clementi_1995} E. Clementi and D.~W.~M. Hofmann, Theochem
\textbf{330}, 17 (1995).

\bibitem{korobov_2002} See, for example, V. I. Korobov, Phys.\ Rev.\ A
  \textbf{66}, 024501 (2002), and references therein.

\bibitem{williamson_2001} A.~J.~Williamson, R.~Q.~Hood, and
  J.~C.~Grossman, Phys.\ Rev.\ Lett.\ \textbf{87}, 246406 (2001).
  

\end{thebibliography}
\end{document}